\documentclass[prl,twocolumn]{revtex4}
\usepackage{dcolumn}
\usepackage{multirow}
\usepackage{amsmath,amsthm,amssymb}
\usepackage{hyperref}
\usepackage{xcolor}

\def\asc{\mathpzc{a}}

\def\fsc{\mathpzc{f}}
\def\qsc{\mathpzc{q}}
\def\psc{p}
\def\msc{\mathpzc{m}}
\def\nsc{\mathpzc{n}}
\def\dpscl{\delta p_\ell}
\def\drhol{\delta \rho_\ell}
\def\dwscl{\delta \mathpzc{w}_\ell}
\def\dvscl{\delta \mathpzc{v}_\ell}
\def\dzscl{\delta \mathpzc{z}_\ell}

\def\Hsc{\mathpzc{H}}

\def\phiM{\tilde\varphi}
\def\cs2{c_s^2}

\newcommand{\hRW}[1]{
h^{(RW)}_{#1}
}

\newcommand{\hRWl}[1]{
h^{(RW)}_{\ell\,\,#1}
}

\DeclareMathAlphabet{\mathpzc}{OT1}{pzc}{m}{it}

\begin{document}
%


\title{
Cosmological perturbations in the Regge-Wheeler formalism}

\author{Andrzej Rostworowski}
\email{arostwor@th.if.uj.edu.pl}
\affiliation{M. Smoluchowski Institute of Physics, Jagiellonian University, 30-348 Krak\'ow, Poland}
%
%
\begin{abstract}
We study linear perturbations of the Friedmann-Lema\^itre-Robertson-Walker (FLRW) cosmological model in the Regge-Wheeler formalism which is a standard framework to study perturbations of spherically-symmetric black holes. In particular, we show that the general solution of linear perturbation equations can be given in terms of two copies of a master scalar satisfying scalar wave equation on the FLRW background (with a Regge-Wheeler/Zerilli type potential) thus representing two gravitational degrees of freedom, and one scalar satisfying a transport type equation representing (conformal) matter perturbation. We expect the Regge-Wheeler formalism to be easily extended to include nonlinear perturbations, akin to to the recent work [Phys. Rev. D 96, 124026 (2017)]. 
\end{abstract}

\maketitle

\paragraph{Introduction.}
\label{Sec:Intro}
While the standard framework for studying cosmological perturbations is based on $1+3$ splitting of metric perturbations resulting in scalar-vector-tensor ($\mathcal{S-V-T}$) sectors of perturbations, the standard framework for studying metric perturbations of the Schwarzschild black hole -- the Regge-Wheeler (RW) formalism \cite{RW} -- is based on $2+2$ splitting of metric perturbation resulting in polar and axial sectors of perturbations (after expansion into suitably chosen polar/axial spherical harmonics). The key result of Schwarzschild black hole perturbation theory is that at linear level the general perturbation can be given in terms of only two (axial/polar) master scalars satisfying scalar wave equation on the Schwarzschild background with Regge-Wheeler \cite{RW} and Zerilli \cite{Zerilli} potentials for axial and polar sectors respectively (more precisely, this holds for any multipole $\ell \geq 2$; the monopole $\ell=0$ and dipole $\ell=1$ cases need some special treatment). Recently, this result was generalized to nonlinear perturbations \cite{r_PRD96}. In this letter, we report that, once the Regge-Wheeler formalism is applied, the same structure emerges also for FLRW perturbations, namely the general solution of linear perturbation equations can be given in terms of two copies of a master scalar satisfying scalar wave equation on the FLRW background (with a Regge-Wheeler/Zerilli type potential) thus representing two gravitational degrees of freedom, and one scalar satisfying a transport type equation representing (conformal) matter perturbations. Our motivation for this study is twofold. First, it is useful to cover both, black hole and cosmological perturbation, with a single framework to allow for simpler exchange of ideas between these two fields of research. Second, we expect that the RW formalism for cosmological perturbations can be extended beyond the linear approximation, as  was shown to be the case for black hole perturbations \cite{r_PRD96}. In general, perturbation approach to Einstein equations consists in trading off the nonlinearity of the equations for the \textit{infinite} system of \textit{linear} perturbation equations. The homogeneous part of these perturbation equations constitutes the linear approximation. The general solution at nonlinear orders is then build from a general solution of the homogeneous part (i.e. a general solution of linear approximation) and some particular solution of the inhomogeneous part of the equations. 
While this letter deals with the linear problem i.e. provides a general solution of linearized Einstein equations in terms of a master scalar, the inhomogeneous part of the problem appearing at higher orders of perturbation expansion is postponed for the future work \cite{rr_preparation}.   

To simplify the notations we assume that perturbations are axially symmetric and we fix the gauge to the RW gauge. At linear level, including azimuthal angle dependence is trivial and at nonlinear orders it would be just a technical, not a conceptual issue. Gauge invariant formulation can be easily provided, but it would make the presentation longer, adding little to the physical content of this letter. 

In fact, the perturbation equations for the FLRW model in RW formalism were derived in \cite{mw_CQG22} and \cite{km_CQG34} for axial and polar sectors respectively, but our treatment of their solution is a novelty, motivated by extension of the RW formalism beyond the linear order \cite{r_PRD96}, which in the case of FLRW perturbations will be the subject of our forthcoming study \cite{rr_preparation}.

\paragraph{Setup.}
\label{Sec:Setup}
We use standard conformal coordinates in which the FLRW line element takes the form:
\begin{equation}
\label{eq:line_element_FLRW}
ds^2 = \asc^2(\tau) \left[ -d\tau^2 + d \qsc^2 + \fsc^2(\qsc) d\Omega^2_2 \right] \, ,
\end{equation}
where $\fsc(\qsc) = \qsc, \, \sin\qsc, \sinh \qsc$ for the flat, closed, and open universes respectively, and $d\Omega^2_2 = du^2/(1-u^2) + (1-u^2) d\phi^2$ is the line element on a unit sphere (with $-1 \leq u = \cos\theta \leq 1$). The perturbed metric takes the form $g_{\mu\nu} = \bar g_{\mu\nu} + \epsilon \hRW{\mu\nu} + \mathcal{O}\left(\epsilon^2 \right)$, where $\bar g_{\mu\nu}$ is FLRW metric, corresponding to the line element (\ref{eq:line_element_FLRW}) and the perturbation $\hRW{\mu\nu}$ is taken in the RW gauge \cite{RW} \footnote{choosing the RW gauge, being uniquely defined, is equivalent to using gauge invariant formulation where one constructs two gauge invariants out of three metric perturbations in the axial sector, and four gauge invariants out of seven metric perturbations in the polar sector}. At axial symmetry, the RW gauge is uniquely defined by the conditions: $\hRW{u\phi} \,=\, 0 \,=\, \hRW{\tau u} \,=\, \hRW{\qsc u} \,=\, (1-u^2)\hRW{uu}-\hRW{\phi\phi}/(1-u^2) =:\hRW{-}$ (in general: the axial component of  $\hRW{u\phi}$ and polar components of $\hRW{\tau u}$, $\hRW{\qsc u}$ and $\hRW{-}$ are zero). The remaining two axial and four polar components of metric perturbations are expanded into multipoles:
\begin{align}
\label{eq:hRWaxial}
\hRW{a\phi}(\tau,\qsc,u) &= \sum_{1 \geq \ell} \hRWl{a\phi}(\tau,\qsc) \left( 1- u^2 \right) P'_\ell(u), \,\, a=\tau,\,\qsc \,,
\\
\hRW{ab}(\tau,\qsc,u) &= \sum_{0 \geq \ell} \hRWl{ab}(\tau,\qsc) P_\ell(u), \qquad a, \, b=\tau, \, \qsc \,,
\end{align}
\begin{align}
\label{eq:hRWpolar+}
&\frac12 \left[ (1-u^2) \hRW{uu}(\tau,\qsc,u) + \frac{1}{1-u^2} \hRW{\phi\phi}(\tau,\qsc,u) \right] 
\nonumber\\
&= \sum_{0 \geq \ell} \hRWl{+}(\tau,\qsc) P_\ell(u) \,,
\end{align}
where $P_\ell$ are Legendre polynomials. 
Since the RW gauge is uniquely defined $\hRW{\tau\phi}$,
$\hRW{\qsc\phi}$ and $\hRW{\tau\tau}$, $\hRW{\tau\qsc}$,
$\hRW{\qsc\qsc}$, $\hRWl{+}$ are in fact two axial and four polar gauge
invariant characteristics of perturbations that we will call RW gauge
invariants (see \cite{nollert_CQG16} for a pedagogical introduction to the RW formalism; technically, due to spherical symmetry of the FLRW background, the expansion (\ref{eq:hRWaxial}-\ref{eq:hRWpolar+}) and (\ref{eq:rho}-\ref{eq:4velocity}) ensures the separation of the angular part of the problem in perturbation Einstein equations (\ref{eq:E-}-\ref{eq:Etu}) and (\ref{eq:Euphi}-\ref{eq:Etphi})). Following \cite{ccf_JCAP06,km_CQG34} we denote polar RW gauge invariants as $\hRWl{+} = \varphi_\ell$, $\hRWl{\qsc\qsc} = (\varphi_\ell + \chi_\ell)$, $\hRWl{\tau\tau} = (\varphi_\ell + \chi_\ell + \psi_\ell)$ and $\hRWl{\tau\qsc} = \sigma_\ell$, thus $\varphi_\ell$ corresponds to \textit{conformal} multipole perturbations \cite{ccf_JCAP06}. To shorten the notation, we also denote $\hRWl{\tau\phi}=\msc_\ell$ and $\hRWl{\qsc\phi}=\nsc_\ell$ in the axial sector ($\varphi_\ell$, $\chi_\ell$, $\psi_\ell$, $\sigma_\ell$, $\msc_\ell$ and $\nsc_\ell$ are functions of $\tau$ and $\qsc$). 

We model the matter content of the universe as a single component perfect fluid  with the energy-momentum tensor
\begin{equation}
\label{eq:energy_momentum_perfect_fluid}
T_{\mu\nu} = (\rho + \psc) u_\mu u_\nu + \psc g_{\mu\nu}\, ,
\end{equation}
thus we exclude anisotropic stresses from our consideration. Including 
multi-component fluids is again a technical issue that we skip for the clarity of presentation. Density $\rho$, pressure $p$ and 4-velocity $u_\mu$ of the fluid expanded to the first order in the perturbation parameter $\epsilon$ read:
\begin{align}
\label{eq:rho}
\rho(\tau,\qsc,u) &= \rho_0(\tau) + \epsilon \sum_{0 \leq \ell} \drhol(\tau,\qsc)
P_\ell(u) + \mathcal{O} \left( \epsilon^2 \right) \, ,
\\
\label{eq:psc}
\psc(\tau,\qsc,u) &= \psc_0(\tau) + \epsilon \sum_{0 \leq \ell} \dpscl(\tau,\qsc)
P_\ell(u) + \mathcal{O} \left( \epsilon^2 \right) \, ,
\\
u_\qsc(\tau,\qsc,u) &= \epsilon \sum_{0 \leq \ell} \dwscl(\tau,\qsc) P_\ell(u) + \mathcal{O} \left( \epsilon^2 \right) \, ,
\\
u_u(\tau,\qsc,u) &= \epsilon \sum_{0 \leq \ell} \dvscl(\tau,\qsc) P'_\ell(u) + \mathcal{O} \left( \epsilon^2 \right) \, ,
\\
u_\phi(\tau,\qsc,u) &= \epsilon \sum_{0 \leq \ell} \dzscl(\tau,\qsc) \left( 1 - u^2 \right) P'_\ell(u) + \mathcal{O} \left( \epsilon^2 \right) \, ,
\end{align} 
\begin{align}
\label{eq:4velocity}
u_\tau(\tau,\qsc,u) &= -\asc(\tau) + \epsilon \sum_{0 \leq \ell} \frac{\hRWl{\tau\tau}(\tau,\qsc)}{2 \asc(\tau)} P_\ell(u) + \mathcal{O} \left( \epsilon^2 \right) \, ,
\end{align} 
where $u_\mu u^\mu = -1 +\mathcal{O} \left( \epsilon^2 \right)$ holds.
We allow a non-zero value of the cosmological constant $\Lambda$. Then from the Einstein equations (we use units such that $8 \pi G =1$)
\begin{equation}
\label{eq:Einstein}
E_{\mu\nu}:=R_{\mu\nu} - \frac12 R g_{\mu\nu} + \Lambda g_{\mu\nu} - T_{\mu\nu} =0 
\end{equation}
 we get at the zeroth order in $\epsilon$ (corresponding to homogeneous, isotropic solution of Friedman equations)
\begin{align} 
\label{eq:rho0}
\rho_0 &= \frac{3}{\asc^2} \left( \Hsc^2 + \frac{1- \fsc'{}^2}{\fsc^2} \right) - \Lambda\, ,
\\
\label{eq:psc0}
\psc_0 &= \Lambda - \frac{1}{\asc^2} \left( \Hsc^2 + \frac{1- \fsc'{}^2}{\fsc^2} + 2 \dot \Hsc \right) \, ,
\end{align}
where $\Hsc = \Hsc(\tau) = \left(\partial_{\tau} \asc(\tau)\right) / \asc(\tau)$ is the \textit{conformal} Hubble constant and $\dot{}$ and ${}'$ denote derivatives with respect to $\tau$ and $\qsc$ respectively. From the fact that $\rho_0$ and $\psc_0$ are functions of conformal time $\tau$ only, it follows that $\left( 1- \fsc'{}^2 \right)/\fsc^2$ must be a constant (which is indeed the case for $\fsc(\qsc) = \qsc, \, \sin\qsc, \sinh \qsc$) thus, from $d\left[\left( 1- \fsc'{}^2 \right)/\fsc^2\right] / d\qsc =0$, it follows
\begin{equation}
\label{eq:fsc2}
\fsc'' = -\left( 1- \fsc'{}^2 \right)/\fsc \, . 
\end{equation}
The system (\ref{eq:rho0},\ref{eq:psc0}) is closed by the equation of state of the fluid $\psc=\psc(\rho)$. Differentiating  (\ref{eq:rho0},\ref{eq:psc0}) with respect to $\tau$ and defining the speed of sound as 
\begin{equation}
\label{eq:cs2}
\cs2 = \left. d\psc / d \rho \right|_{\rho=\rho_0} \, ,
\end{equation} 
we get
\begin{equation}
\label{eq:Hsc2}
\ddot \Hsc = \left( 1+ 3 \cs2 \right) \Hsc \left( \Hsc^2 + \frac{1- \fsc'{}^2}{\fsc^2} \right) + \left( 1- 3 \cs2 \right) \Hsc \dot \Hsc \, .
\end{equation}
In the following, we use (\ref{eq:fsc2}, \ref{eq:Hsc2}) to get rid of derivatives of $\fsc$ and $\Hsc$ higher than the 1st order. 


\paragraph{Linear perturbations.}
\label{Sec:LinearPerturbation}

We start with perturbations in the polar sector. In the polar sector, at linear order in $\epsilon$, (\ref{eq:Einstein}) yield \footnote{we define $\pm$ components of Einstein equations as $E_\pm = (1-u^2) E_{uu} \pm E_{\phi\phi}/(1-u^2)$}:
\begin{align}
\label{eq:E-}
E_{-}:\,\, & \psi_\ell = 0\, ,
\\
\label{eq:Equ}
E_{\qsc u}:\,\, & \chi_{\ell}' - \dot \sigma_{\ell} = 0 \, ,
\\
\label{eq:E+}
E_{+}:\,\, & \ddot \chi_{\ell} - \chi_{\ell}'' + \frac{2 \fsc'}{\fsc} \chi_{\ell}'  - 2 \left(\Hsc \chi_{\ell}\right) \dot{}  + \frac{\ell(\ell+1) - 2}{\fsc^2} \chi_{\ell} = 0 \, ,
\end{align}
\begin{widetext}
\begin{align}
\label{eq:Ett}
E_{\tau \tau}:\,\, &  2 \asc^4 \drhol = \left( \frac{\ell(\ell+1) + 6 \fsc'^2 - 4}{\fsc^2} + 2\Hsc^2\right) \chi_\ell + 2 \left( \frac{\ell(\ell+1) + 3 \fsc'^2 - 3}{\fsc^2} - 3\Hsc^2\right) \varphi_\ell
\nonumber\\
& + 2 \frac{\fsc'}{\fsc} \left( \chi_\ell' + 2 \varphi_\ell' -4 \Hsc \sigma_\ell \right) + 2 \Hsc \left( \dot \chi_\ell + 3 \dot \varphi_\ell - 2 \sigma_\ell' \right) - 2 \varphi_\ell'' \, ,
\\
\label{eq:Eqq}
E_{\qsc \qsc}:\,\, & 2 \asc^4 \dpscl = \left( \frac{\ell(\ell+1) - 2 \fsc'^2}{\fsc^2} + 2\Hsc^2 - 4 \dot \Hsc \right) \chi_\ell + 2 \left( \frac{1 - \fsc'^2}{\fsc^2} + \Hsc^2 \right) \varphi_\ell  + 2 \frac{\fsc'}{\fsc} \chi_\ell' + 2 \Hsc \left( \dot \varphi_\ell - \dot \chi_\ell \right) - 2 \ddot \varphi_\ell \, ,
\\
\label{eq:Etq}
E_{\tau \qsc}:\,\, & 2 \asc^3 \left( \rho_0 + \psc_0 \right) \dwscl = - \frac{\ell(\ell+1) + 4 \fsc'^2 - 4}{\fsc^2} \sigma_\ell + 2 \frac{\fsc'}{\fsc} \left( 2 \Hsc \chi_\ell - \dot \chi_\ell \right) + 2 \Hsc \left( \chi_\ell' - \varphi_\ell' \right) + 2 \dot \varphi_\ell' \, ,
\\
\label{eq:Etu}
E_{\tau u}:\,\, & 2 \asc^3 \left( \rho_0 + \psc_0 \right) \dvscl = -2 \Hsc \varphi_\ell - \sigma_\ell' + 2 \dot \varphi_\ell + \dot \chi_\ell \, .
\end{align}
In fact, equations (\ref{eq:Ett}-\ref{eq:Etu}) give perturbations of the material content of the universe in terms of metric perturbations. Using (\ref{eq:cs2}) we combine (\ref{eq:Ett},\ref{eq:Eqq}) to yield
\begin{align}
\label{eq:phi}
&-\ddot \varphi_\ell 
+ \cs2 \varphi''_\ell 
+ 2 \cs2 \frac{\fsc'}{\fsc} \varphi'_\ell 
+ \left( 1- 3 \cs2 \right) \Hsc \dot \varphi_\ell
+ \frac{\left( 1+ 3 \cs2 \right) \left( 1 + \fsc^2 \Hsc^2 - \fsc'^2  \right) - \cs2 \ell(\ell+1)}{\fsc^2} \varphi_\ell 
\nonumber\\
&= \left[ \frac{ \left( 1+ 3 \cs2 \right) \fsc'^2 - 2}{\fsc^2} + \left( \cs2 - 1 \right) \left( \frac{\ell(\ell+1)}{2 \fsc^2} + \Hsc^2 \right) + 2 \dot \Hsc \right] \chi_\ell - 2 \cs2 \Hsc \left( \sigma_\ell' + 2 \frac{\fsc'}{\fsc} \sigma_\ell \right) + \frac{\fsc'}{\fsc} \left( \cs2 - 1 \right) \chi_\ell' + \Hsc \left( \cs2 + 1 \right) \dot \chi_\ell \, .
\end{align}
\end{widetext}
In fact, the system (\ref{eq:Equ},\ref{eq:E+},\ref{eq:phi}) can be written as a system of four, first order in time equations, thus its solution is entirely specified by four profiles representing initial data. The equation (\ref{eq:phi}) is inhomogeneous equation for $\varphi_\ell$ thus the simplest solution of the system (\ref{eq:Equ},\ref{eq:E+},\ref{eq:phi}) is given by $\chi_\ell = 0 = \sigma_\ell$ and $\varphi_\ell =\phiM_\ell$ with $\phiM_\ell$ being the solution of the homogeneous part of (\ref{eq:phi}). 
This solution represents \textit{conformal} material perturbations. For the dust case ($\cs2=0$), the left hand side of eq. (\ref{eq:phi}) reduces to 
\begin{equation}
-\asc^2 \left[\frac{1}{\asc} \partial_\tau \left( \frac{1}{\asc} \partial_\tau \varphi_\ell \right) - \left( \Hsc^2 + \frac{1-\fsc'^2}{\fsc^2}\right) \frac{\varphi_\ell}{\asc^2}\right] 
\end{equation}
thus, we call (\ref{eq:phi}) a transport equation. To find the general solution including gravitational waves we resort to the guiding principle introduced in \cite{r_PRD96}, namely we assume that $\sigma_\ell$, $\chi_\ell$ and $\phi_\ell$ are given in terms of linear combinations of a master scalar $\Phi_\ell=\Phi_\ell(\tau,\qsc)$ and its derivatives with their coefficients being $(\tau,\qsc)$ dependent functions, where the master scalar itself satisfies a scalar wave equation on FLRW background with a potential:
\begin{equation} 
\label{eq:wave}
\left( -\Box_{\bar g} + V_\ell \right) \frac{\Phi_\ell}{\fsc} = 0 \,,
\end{equation} 
thus representing one (polar) gravitational degree of freedom. 
The general strategy is to try to include the lowest order derivatives of $\Phi_\ell$ in these linear combinations, and if this results in a trivial solution, to include higher derivatives until ultimately a non-trivial solution is found. Plugging such ansatz into the set of equations to be solved \footnote{Although for the perfect fluid (\ref{eq:energy_momentum_perfect_fluid}) the equations of motion of the fluid, $\nabla_\mu T^{\mu\nu} = 0$, are satisfied identically once the Einstein equations are fulfilled, we tactically include their polar/axial components into the set of equations we are trying to solve with our ansatz to get more information on the algebraic relations between the coefficients of linear combinations of $\Phi_\ell$ and its derivatives in the ansatz.} we start with equating to zero the coefficients of the highest derivatives of $\Phi_\ell$ present in the resulting expressions. This, at the beginning, results in some set of algebraic relations between the coefficients in the ansatz, and ultimately we end up with a system of partial differential equation for a single coefficient that is easy to solve; its solution introduces one (multiplicative) integration constant as should be expected for the solution of an homogeneous linear system (also the solution of the (homogeneous) wave equation (\ref{eq:wave}) is given up to a multiplicative constant). All other non-vanishing coefficients depend linearly on this solution and the potential in the wave equation (\ref{eq:wave}) is also set as a consistency condition. Indeed, proceeding along these lines, we get the uniquely defined general solution of the system (\ref{eq:Equ},\ref{eq:E+},\ref{eq:phi}) in  the form:
\begin{align}
\label{eq:sigmaSol}
\sigma_\ell &= \asc^2 \left[ \fsc \Phi''_\ell - \frac{\ell(\ell+1)}{\fsc} \Phi_\ell \right]'\, ,
\\
\label{eq:chiSol}
\chi_\ell &= \asc^2 \left[ \fsc\left( \dot \Phi''_\ell + 2 \Hsc \Phi''_\ell \right) - \frac{\ell(\ell+1)}{\fsc} \left( \dot \Phi_\ell + 2 \Hsc \Phi_\ell \right) \right] \, ,
\\
\label{eq:phiSol}
\varphi_\ell &= \asc^2 \left[ \fsc' \left( \dot \Phi'_\ell + \Hsc \Phi'_\ell \right) - \fsc \Hsc \Phi''_\ell
+ \frac{\ell(\ell+1)}{2 \fsc} \left( \dot \Phi_\ell + 3 \Hsc \Phi_\ell \right)\right] 
\nonumber\\
& + \phiM_\ell \, ,
\end{align}
where $\Phi_\ell$ solves (\ref{eq:wave}) with
\begin{equation}
\label{eq:Vpolar}
V_\ell = V_\ell (\tau,\qsc) = \frac{1}{\asc^2} \left( \frac{\ell(\ell+1)}{\fsc^2} + 2 \dot \Hsc \right) \, .
\end{equation}
This potential is an analogue of the Zerilli potential \cite{Zerilli} in the polar sector of Schwarzschild black hole perturbations.


For the sake of completeness, we discuss the axial sector of perturbations. In the axial sector, at linear order in $\epsilon$, (\ref{eq:Einstein}) yield:
\begin{align}
\label{eq:Euphi}
E_{u \phi}:\,\, & \nsc_\ell' - \dot \msc_\ell  = 0\, ,
\\
\label{eq:Eqphi}
E_{\qsc \phi}:\,\, & \ddot \nsc_{\ell} - \dot \msc_{\ell}' + 2 \frac{\fsc'}{\fsc} \dot \msc - 2 \left(\Hsc \nsc_{\ell}\right) \dot{} + \frac{\ell(\ell+1) - 2}{\fsc^2} \nsc_{\ell} = 0 \, ,
\\
\label{eq:Etphi}
E_{\tau \phi}:\,\, & 2 \asc^3 \left( \rho_0 + \psc_0 \right) \dzscl = - \frac{\ell(\ell+1) + 4 \fsc'^2 - 4}{\fsc^2} \msc_\ell 
\nonumber\\
& \hskip10mm+ 2 \frac{\fsc'}{\fsc} \left( 2 \Hsc \nsc_\ell - \dot \nsc_\ell \right) + 2 \Hsc \nsc_\ell'  - \dot \nsc_\ell' + \msc_\ell''\, .
\end{align}
Again, eq. (\ref{eq:Etphi}) gives an axial perturbation of the material content of the universe in terms of the axial metric perturbations. Is is straightforward (as was done in \cite{mw_CQG22, km_CQG34}), to combine (\ref{eq:Euphi},\ref{eq:Eqphi}) to yield a single scalar wave-type equation of the form (\ref{eq:E+}) (with a substitution $\chi_\ell \rightarrow \nsc_\ell$) but we prefer to give a clear geometrical meaning to the axial master scalar, as was the case in the polar sector. To this end we again make the ansatz that $\msc_\ell$ and $\nsc_\ell$ in (\ref{eq:Euphi},\ref{eq:Eqphi}) are given in terms of linear combinations of a master scalar $\Phi_\ell$ and its derivatives, where the master scalar satisfies the scalar wave equation (\ref{eq:wave}) (with possibly some different potential than the polar one (\ref{eq:Vpolar})). As the result of such substitution we get
\begin{align}
\label{eq:mscSol}
\msc_\ell &= \left( \asc^2 \fsc \, \Phi_\ell \right)'\, ,
\\
\label{eq:nscSol}
\nsc_\ell &= \left( \asc^2 \fsc \, \Phi_\ell \right)\dot{} \, ,
\end{align}
with the same potential in (\ref{eq:wave}) as in the polar sector, i.e. the one given in (\ref{eq:Vpolar}). Thus, in the case of FLRW perturbations, the analogues of the Regge-Wheeler \cite{RW} and Zerilli \cite{Zerilli} potentials have the same form, both in polar/axial sectors, given in (\ref{eq:Vpolar}). Then, $\dzscl$, as given by (\ref{eq:Etphi}), is identically zero after substitution of (\ref{eq:mscSol},\ref{eq:nscSol}) (for sufficiently smooth solutions of (\ref{eq:wave})), as expected. 

Once the Einstein equations  (\ref{eq:E-}-\ref{eq:Etu},\ref{eq:Euphi}-\ref{eq:Etphi}) (together with (\ref{eq:phi})) are satisfied then the equations of motion of the fluid, $\nabla_\mu T^{\mu\nu} = 0$, are identically satisfied for a perfect fluid (\ref{eq:energy_momentum_perfect_fluid}). 


To summarize, to satisfy the set of linear perturbation Einstein equations for the FLRW model (\ref{eq:E-}-\ref{eq:phi}, \ref{eq:Euphi}-\ref{eq:Etphi}) it is enough to solve just one master scalar equation (\ref{eq:wave}) with a potential given by (\ref{eq:Vpolar}) for the gravitational master scalar $\Phi_\ell$ and (in the polar sector) to find $\phiM_\ell$, solving the homogeneous part of the transport equation (\ref{eq:phi}). Then, in the polar sector, the metric perturbations are given by (\ref{eq:E-},\ref{eq:sigmaSol},\ref{eq:chiSol},\ref{eq:phiSol}) and the material perturbations can be read off directly from Einstein equations (\ref{eq:Ett}-\ref{eq:Etu}). In the axial sector, the metric perturbations are given by (\ref{eq:mscSol},\ref{eq:nscSol}) and the material perturbation can be read off from (\ref{eq:Etphi}). The solution for $\Phi_\ell$ and $\phiM_\ell$ are given in terms of initial data for (\ref{eq:wave}) and (\ref{eq:phi}) i.e. the profiles $\left. \Phi_\ell \right|_{\tau_0}$, $\dot \Phi_\ell |_{\tau_0}$, $\left. \phiM_\ell \right|_{\tau_0}$, $\left. \dot \phiM_\ell \right|_{\tau_0}$ at some initial conformal time slice $\tau_0$. Of course, the initial profiles for the master scalar $\left. \Phi_\ell \right|_{\tau_0}$, $\dot \Phi_\ell |_{\tau_0}$ are to be set independently for polar/axial sectors of perturbations.

\paragraph{de Sitter limit.}
\label{Sec:deSitterLimit}
To satisfy the perturbation Einstein equations (\ref{eq:E-}-\ref{eq:Etu},\ref{eq:Euphi}-\ref{eq:Etphi}) in the de Sitter limit ($\rho_0=0=\psc_0$ in (\ref{eq:rho},\ref{eq:psc},\ref{eq:rho0},\ref{eq:psc0})) one can use (\ref{eq:sigmaSol}-\ref{eq:phiSol},\ref{eq:mscSol},\ref{eq:nscSol}) with (\ref{eq:wave},\ref{eq:Vpolar}) again, with the only difference that now, as a result of (\ref{eq:Etu}), we have $\dot \phiM_\ell = \Hsc \phiM_\ell$, instead of the homogeneous part of (\ref{eq:phi}). Interestingly, in this limit, $\dpscl$, as given by (\ref{eq:Eqq}), is identically zero. In fact, in the de Sitter limit, instead of (\ref{eq:sigmaSol}-\ref{eq:phiSol}), one can also use simpler expressions (that do not contain 3rd derivatives of the master scalar):
\begin{align}
\label{eq:sigmaSoldeSitter}
\sigma_\ell &= \asc^2 \left[ \fsc \left( \dot \Phi_\ell + \Hsc \Phi_\ell \right) \right]'\, ,
\\
\label{eq:chiSoldeSitter}
\chi_\ell &= \asc^2 \left[ \fsc \left( \Phi''_\ell + \Hsc \dot\Phi_\ell + \Hsc^2 \Phi_\ell \right) - \frac{\ell(\ell+1)}{\fsc} \Phi_\ell \right] \, ,
\\
\label{eq:phiSoldeSitter}
\varphi_\ell &= \asc^2 \left[ \fsc' \Phi'_\ell - \fsc \left( \Hsc \dot \Phi_\ell + \Hsc^2 \Phi_\ell \right) + \frac{\ell(\ell+1)}{2 \fsc} \Phi_\ell \right] + \phiM_\ell \, ,
\end{align}
where again $\Phi_\ell$ solves (\ref{eq:wave}) with (\ref{eq:Vpolar}). Formulas (\ref{eq:sigmaSoldeSitter}-\ref{eq:phiSoldeSitter}) can be transformed into (\ref{eq:sigmaSol}-\ref{eq:phiSol}) with the substitution $\Phi_\ell \rightarrow \tilde \Phi_\ell = \Hsc \Phi_\ell + \dot \Phi_\ell$ since if $\Phi_\ell$ solves (\ref{eq:wave}) with (\ref{eq:Vpolar}), so does $\tilde \Phi_\ell$ in the de Sitter case ($\rho_0=0=\psc_0$).

\paragraph{Conclusions.}
\label{Sec:Conclsions}
We have shown that the general solution of linear perturbations of FLRW cosmological model can be reduced to the problem of solving just one master scalar wave equation whose two copies govern polar and axial sectors of perturbations (akin to the black holes perturbations), together with finding the solution to the homogeneous part of the transport equation (\ref{eq:phi}). This is a great simplification since the scalar wave equation on FLRW background can be easily treated, at least numerically (see also \cite{klainerman_sarnak}). Moreover our  framework should be easily extended to include nonlinear cosmological perturbations, akin to our recent scheme \cite{r_PRD96}. This is the subject of our ongoing work \cite{rr_preparation}. Of course, the cosmological perturbations given in the Regge-Wheeler formalism can be easily translated into the standard language of ($\mathcal{S-V-T}$) perturbations as discussed in the Appendix D of \cite{ccf_JCAP06}, but it may be advantageous, in view of comparison with observations, that the Regge-Wheeler framework operates directly with spherical multipoles. 

We find it rather remarkable that gravitational perturbations of exact solutions of Einstein equations are in fact governed by two scalar functions, satisfying scalar wave equations (with a potential) on the background solution, corresponding to two polarizations of gravitational waves. In our opinion a deeper understanding of this fact is an interesting mathematical problem.  

\paragraph{Acknowledgments.}
 This research was supported by the Narodowe Centrum Nauki (Republic of Poland) Grant no. 2017/26/A/ST2/530. The author is indebted to Piotr Bizo\'n for constant peaceful support in his research and for careful reading of the early version of the manuscript.

\end{document}